\documentclass[pra,twocolumn,showpacs,amsmath,floatfix]{revtex4-1}
\usepackage{graphicx}
\usepackage{dcolumn}
\begin{document}
\preprint{}

\title{Superparabolic Level Glancing Models for Two-State Quantum Systems}
\author{J. Lehto}
\email{Jaakko.Lehto@utu.fi}
\author{K.-A. Suominen}
\email{Kalle-Antti.Suominen@utu.fi}
\affiliation{Turku Centre for Quantum Physics, Department of Physics and Astronomy, University of Turku, FI-20014 Turku, Finland}
\date{\today}
\begin{abstract}
Level crossing models for two-state quantum systems are applicable to a wide variety of physical problems. We address the special case of level glancing, i.e., when energy levels reach a degeneracy at a specific point of time, but never actually cross. The simplest model with such behaviour is the parabolic model, and its generalizations, which we call superparabolic models. We discuss their basic characteristics, complementing the previous work on the related nonlinear crossing models [Phys. Rev. A \textbf{59}, 4580 (1999)]. 
\end{abstract}

\maketitle

\section{Introduction}

The level crossing models provide an important tool for the description of nonadiabatic transitions. These transitions occur when the energy levels of coupled quantum states are brought close to each other by driving them with external fields. The most prominent example of these models, the Landau-Zener (LZ) model, was introduced already in 1932~\cite{landau,zener,Stuckelberg1932} in connection of atomic collisions (and by Majorana for atoms in time-varying magnetic fields~\cite{Majorana1932}), but because time-dependent level crossing problems have been studied over the years in connection with a wide variety of phenomena in physics and chemistry, for example with laser-atom interactions \cite{kazantsev,metcalf,Shore2011}, laser-induced molecular dynamics~\cite{Garraway1995}, slow~\cite{Nikitin1984,Child1991,Nakamura2012} and cold~\cite{suominen1} atomic collisions, molecular collisions~\cite{child1}, neutrino oscillations~\cite{parke,keranen} and in Bose-Einstein condensation \cite{jona-lasinio}, including evaporative cooling of atoms~\cite{Suominen1998}, outcouplers for atom lasers~\cite{Mewes1997,Vitanov1997,Martikainen1999,Bloch2001} and association of cold atoms into molecules~\cite{Ishkhanyan2004,Ishkhanyan2009}.  In the recent years they have raised a lot of interest especially in nanophysics and quantum information processing as a way to coherently control qubits~\cite{kayanuma} and demonstrations of this have been done for example with solid-state artificial atoms \cite{sillanpaa, oliver}. Lastly, Landau-Zener type models have played an important role as a tool to understand the Kibble-Zurek theory of topological defect production and the dynamics of the quantum phase transitions~\cite{damski,barankov,Dziarmaga2010,Polkovnikov2011}. 

First we define some terminology for a quantum system with two coupled states and explicit time dependence in the Hamiltonian. The actual \textit{level crossing} happens in the diabatic basis which consists of the bare states, i.e., system eigenstates when no coupling is present. The energy difference between the diabatic states is termed as detuning, which becomes zero at the degeneracy point. However, such crossing appears only as an \textit{avoided crossing} in the adiabatic basis (the basis of the instantaneous eigenstates of the time-dependent Hamiltonian) due to the coupling between the diabatic states. If the levels never reach degeneracy in the diabatic basis but get close doing so at some instant of time, transitions can happen by tunnelling. In this paper we concentrate on a third kind of situation, which occurs as a limiting case between a proper crossing and a tunneling case, namely that the diabatic levels merely touch each other momentarily but do not actually cross. We call this a \textit{level glancing}. 

In the original Landau-Zener model the diabatic energy levels change linearly in time and the diabatic coupling is constant. Although this is a very crude assumption it has been applied very successfully over the years. The reason for this is that the nonadiabatic transition is located in the vicinity of the crossing point and one can usually linearize the diabatic levels in its neighbourhood while the coupling does not vary much during this interval. In the recent years, however, there has been a growing interest to study more general dynamics than the one given by the single-crossing LZ case, for example in different interferometric schemes~\cite{gaitan, shevchenko}. One should note that in the original atomic collisions problem with semiclassical trajectories, and which motivated Landau~\cite{landau}, Zener~\cite{zener} and St{\"u}ckelberg~\cite{Stuckelberg1932} in 1932, the level crossing was traversed twice, and the phase difference accumulated between crossings gives oscillations to the transition probabilities named after St{\"u}ckelberg, and they have been discussed and observed well before the recent interest in nanosystems, see e.g.~\cite{Yoakum1992}. In this article we follow the practice of refering to the single crossing case as the LZ model, and the "double-crossing" model with phase-related oscillations in transition probabilities as the LZS model. In collision physics the limit where the two crossing points in the LZS model approach each other relates to the case where the classical turning point overlaps with the region of degeneracy, and is accordingly considered as a breakdown of the LZS description. However, that limit is also an example of a level glancing situation. Sadly, for historical reasons, the already more or less standard naming convention does not acknowledge Majorana's contribution~\cite{Majorana1932}.

Another example of more general models are those with cubic-like detuning, i.e. the detuning is proportional an odd power of time. Such models have been analysed previously in Ref.~\cite{vitanov}, and they have recently raised some interest in the study of quantum phase transitions and adiabatic quantum computation~\cite{barankov}. In such cases one typically has to optimize between the computational time and the density of defects produced when crossing the critical point, so dynamics different from the one given by the LZ model is needed, since in the adiabatic limit the transition time for the original LZ model increases exponentially with the coupling strength~\cite{Vitanov1999}. The cubic-like models deviate both quantitatively and qualitatively from the LZ model and were therefore dubbed essentially nonlinear crossing models in Ref.~\cite{vitanov}. 

In this paper we consider the level glancing models that are reminiscent of the LZ and cubic-like models, having nonlinear parabolic-like time-dependent detunings and study the effect of these different time dependencies on the transition probability. One motivation behind this current work is the previously studied time-dependent parabolic model~\cite{suominen2,shimshoni,Teranishi1997}. It can be used to describe situations outside the scope of the usual LZ model, and it encompasses within a single model the cases of tunnelling, double crossings and level glancing dynamics. Such parabolic time dependence of the diabatic energies has been recently applied e.g. in studies of laser-induced molecular dynamics~\cite{Zou2005,Chang2010}. The parabolic model has a peculiar property: in the level glancing case the maximum probability of a nonadiabatic transition is only a little bit over one half~\cite{suominen2}. Unlike in the LZ and cubic-like models, it does not reach unity in the adiabatic limit (actually, it goes to zero for the rather obvious reasons discussed later). Another important aspect is that the transition probabilities show oscillatory behavior for the level glancing situation as a function of relevant parameters, which is expected for any model with double crossings, but which are absent for the LZ model, for instance.

We examine how the energy level dynamics affect the above properties by considering a set of level glancing models with \textit{superparabolic} time dependence, i.e. the detuning depends on some even power of time.  The oscillatory character of the parameter dependence of the final transition probability can be understood on the basis of the structure of the complex zeros of the adiabatic eigenenergies as explained by the Dykhne-Davis-Pechukas (DDP) theory~\cite{dykhne,dp,Suominen1991}. The results of the DDP theory are asymptotically exact in the adiabatic limit and although the structure of the zero points are rather similar in both the cubic-like and superparabolic models, the diabatic limits of these models are completely different as mentioned before. 

The outline of the paper is as follows. In Sec.~\ref{sec:formalism} the basic equations and definitions are given. We also discuss the time evolution of the transition probabilities as well as complementary analytical approximation methods for the final transition probabilities, namely the above-mentioned Dykhne-Davis-Pechukas theory and various perturbation methods. In Sec.~\ref{sec:results} we present and analyze the results that were obtained by numerical calculations and compare these to the approximative expressions. Finally, the discussion in Sec.~\ref{sec:conclusions} ends the presentation.

\section{Formalism}
\label{sec:formalism}
\subsection{Basic equations}
We study the time evolution given by the Schr{\"o}dinger equation $(\hbar = 1)$
\begin{equation}
\dot{\imath} \dfrac{d}{dt}\psi(t) = H(t) \psi(t),
\label{eqn:scheq}
\end{equation}
where $H(t)$ is the general Hamiltonian of the two-level system with real-valued detuning $\Delta(t)$ and coupling $\Omega(t)$, given explicitly by
\begin{equation}
H(t) = \begin{pmatrix} \Delta (t) & \Omega(t) \\
\Omega(t)  & - \Delta (t)\end{pmatrix}
\label{eqn:hamiltonian}					   
\end{equation}
and $\psi(t) = \left[ c_{1}(t), c_{2}(t) \right]^{T} $, where $c_{1}(t)$ and $c_{2}(t)$ are the probability amplitudes of the diabatic basis states $\psi_{1}$ and $\psi_{2}$, respectively. As already mentioned, the level crossing in the diabatic basis is converted to an avoided one in the adiabatic basis, assuming that the coupling $\Omega(t)$ does not vanish at the crossing. The adiabatic basis is formed by the instantaneous eigenstates $ \chi_{1}(t)$ and $\chi_{2}(t)$. The eigenvalues $ \pm \mathcal{E}(t)$ of the Hamiltonian in Eq.~(\ref{eqn:hamiltonian}) form now the adiabatic levels, with
\begin{equation}
\mathcal{E} (t) = \sqrt{\Omega(t)^{2} + \Delta^{2}(t)},
\end{equation}
while the nonadiabatic coupling between the adiabatic states is
\begin{align}
\gamma (t)& \equiv \langle \chi_{1}(t) \mid \dot{\chi}_{2}(t) \rangle  = - \langle \chi_{2}(t) \mid \dot{\chi}_{1}(t) \rangle \nonumber \\
		  & = \pm \frac{ \Omega(t) \dot{\Delta}(t) - \Delta(t) \dot{\Omega}(t) }{2 \left( \Delta(t)^{2} + \Omega(t)^{2}\right)},
\label{eqn:nonadiabcoupl}
\end{align}
where the overhead dot stands for time derivative and one can fix the sign by fixing the relative sign of the basis vectors.

\subsection{Models}

\begin{figure}[htb]
\begin{center}
\includegraphics[scale=0.8]{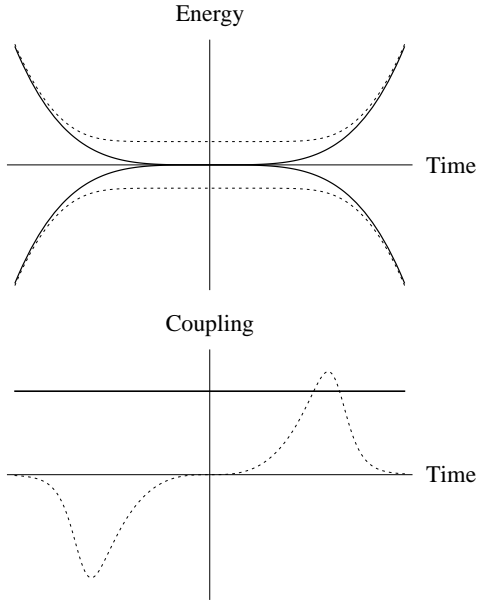} 
\caption{The time-dependence of the energy levels for the superparabolic model with $N = 4$ and $\beta = \Omega = 1$. The diabatic levels and the corresponding coupling are drawn with solid lines while adiabatic ones are drawn with dashed lines.}
\label{tasot}
\end{center}
\end{figure}

We consider models where the coupling $\Omega(t) $ is constant and the detuning $\Delta (t)$ is directly proportional to some even power of time,
\begin{equation}
\Omega (t) = const. \equiv \Omega_{0}, \quad \Delta (t) = \beta^{N + 1} t^{N},
\label{eqn:superparabolic}
\end{equation} 
where $\beta$ and $\Omega_{0}$ are real numbers and are both chosen to be positive and $N = 2, 4, 6, \ldots$ is an even integer number. Now, instead of a crossing, the diabatic levels only touch each other at the point $t = 0$ so we have a level glancing. An example of the energy level structures and couplings is depicted in Fig.~\ref{tasot}. These are fairly similar to other values of $N$ as well. As the $N$ increases, the parabolic-shaped energy levels transform to more rectangular ones and the two peaks of the adiabatic coupling get sharper. The corresponding physical picture is of course that we drive the system fast to the resonance by changing the frequency of the constant-amplitude driving field, keep it there some time, and then the system is brought out of the resonance in a symmetrical fashion.

The case with $N = 2$ is called the parabolic model and it has been introduced in the context of atomic collisions~\cite{Nikitin1984,Child1991,crothers} and later used to study especially coherence effects related to multiple crossings and in situations where the LZ linearization fails~\cite{shimshoni, suominen2,Teranishi1997}. Usually the diabatic level energies in the parabolic model are given by $a t^{2} - b $ so that the level glancing model (\ref{eqn:superparabolic}) is actually only a special case of this with $b = 0$. The parabolic level glancing model is able to give dynamics quite different from the LZ model and is experimentally feasible, but one can not obtain with any parameters a higher probability of nonadiabatic transition than just over one half. We refer to the models with $N = 4, 6, 8, \ldots$ as superparabolic and show that with these models the situation is improved. The models excluded from here where $N$ is an odd integer number have been previously studied in Ref.~\cite{vitanov}. Note also that the case with $N = 1$ is the exactly solvable LZ model. No exact solutions in a sufficiently simple or closed form seem to exist for the parabolic or superparabolic models, nor for any model for which the dynamics of the energy levels are even approximately given by Eq.~(\ref{eqn:superparabolic}) near $t=0$. This is in contrast to e.g. the Demkov-Kunike models with $\Delta(t)\propto \tanh(\beta t)$; they are exactly solvable and reduce to the LZ model with certain parameter limits~\cite{Shore2011,Suominen1992}. The models can be solved exactly, though, using the Zhu-Nakamura method based on Stokes parameters~\cite{Nakamura2012,Zhu1992a,Zhu1992b}, but the results are quite complicated, involving series solutions that eventually have to be evaluated using approximations and numerics (see, e.g., discussion in Ref.~\cite{Osherov2010}). Our motivation is to find reasonably accurate solutions, which have sufficiently simple expressions, and to examine in detail the DDP approximation.

We take the system to be initially in state 2 so that the initial conditions are $c_{2}(-\infty) = 1$ and $c_{1}(-\infty) = 0$ and we are interested in the asymptotic transition probability $P$. In the diabatic basis this is given by $P=\vert c_{1}(\infty) \vert^{2}$. However, as the diabatic and adiabatic bases differ only by a phase factor as $t \rightarrow \pm \infty$ and, unlike in the odd-$N$ cases, the basis states do not swap their labels this is the same as the probability $P$ of a nonadiabatic transition. It should be also noted that now both the adiabatic and diabatic/sudden approximations correspond to $P \approx 0$.

\subsection{Time evolution of populations}

Before discussing the final transition probabilities and their parameter dependence, we study the characteristics of the real time evolution of the populations by transforming to the rotating frame and using the Bloch sphere representation. The change of basis to the rotating frame is obtained by the unitary transformation $U(t) = \exp \left[ - \dot{\imath} \phi \left( t \right) \hat{\sigma}_{z} \right]$. Defining $\tilde{\psi}(t) = U(t) \psi (t)$ the Schr{\"o}dinger equation (\ref{eqn:scheq}) in this new basis reads 
\begin{equation}
\dot{\imath} \dfrac{d}{d t}\tilde{\psi}(t) = \begin{pmatrix} \Delta (t) -  \dot{\phi}(t) & \Omega(t)e^{-2\dot{\imath} \phi(t)} \\
\Omega(t)e^{+2\dot{\imath} \phi(t)}  & - \left(\Delta (t) -  \dot{\phi}(t) \right) \end{pmatrix} \tilde{\psi}(t).
\label{eqn:windingscheq}
\end{equation}
The above change of basis shows that the time-dependence of the detuning can arise from the coupling, which is the case e.g.~when coupling atomic or molecular states with chirped laser pulses~\cite{Garraway1995,Chang2010,Paloviita1995}. Generally, if we choose the time-dependence for the transformation to be $ \phi \left( t \right) = \int^{t} \Delta \left( s \right) ds$, 
we will end up with the Hamiltonian
\begin{equation}
\tilde{H}(t) = \begin{pmatrix} 0 & \Omega \left( t \right) e^{-2\dot{\imath}\phi\left(t \right)} \\
\Omega \left( t \right)  e^{+2\dot{\imath}\phi\left(t \right)}  & 0\end{pmatrix}
\label{eqn:windinghamiltonian}					   
\end{equation}
This shows that any detuning can be alternatively mapped into a property of the coupling. Of course, as suggested by the Hamiltonian in Eq.~(\ref{eqn:windingscheq}), there can be time-dependence in both the detuning and the phase of the coupling, as in the so called winding models~\cite{Nakamura1997}.

Now we study the dynamics of the density matrix $\rho$ in this basis given by the Liouville equation
\begin{equation}
\dot{\imath}\dot{\rho}(t) = [\tilde{H}(t), \rho(t) ].
\label{eqn:liouville}
\end{equation}
Furthermore, we define the three-component Bloch vector $\vec{R}(t) = \left[x(t), y(t), z(t) \right]^{T}$, with 
\begin{equation}
\begin{aligned}
x& = \rho_{21} + \rho_{12}, \\
y& = \dot{\imath}\left(\rho_{12} - \rho_{21} \right), \\
z&= \rho_{11} - \rho_{22}.
\end{aligned}
\label{eqn:xyz}
\end{equation}
For closed systems $\rho_{11} + \rho_{22} = 1$ so that the third component of the Bloch vector gives also the evolution of the transition probability $\vert c_{1}(t) \vert^{2} = \frac{1}{2}\left[1 + z(t)\right]$. From Eq.~(\ref{eqn:liouville}) we get the set of differential equations 
\begin{equation}
\begin{aligned}
\dot{x}& = 2\Omega \sin\left(2\phi\right)z, \\
\dot{y}& = -2\Omega \cos\left(2\phi\right)z, \\
\dot{z}&= 2\Omega \cos\left(2\phi\right)y - 2\Omega \sin\left(2\phi\right)x.
\end{aligned}
\label{eqn:xyzderivative}
\end{equation}
Equation~(\ref{eqn:xyzderivative}) is equivalent to $\dot{\vec{R}} = \vec{B}\times\vec{R}$, where the external field vector is given by
\begin{equation}
\vec{B}(\tau) = 2\Omega \left( t \right)[\cos\left(2\phi(t)\right), \sin\left(2\phi(t)\right), 0]^{T}.
\end{equation}
The magnitude of the field vector is twice the value of the diabatic coupling, $\vert \vec{B} \vert = 2\Omega \left( t \right)$, and it is confined in the $x$-$y$ plane with its direction determined by the angle $ 2\phi(t)$. 

In the case of the superparabolic models (\ref{eqn:superparabolic}) studied in this paper, the field vector traces a circle of radius $2\Omega_{0}$ in the $x$-$y$ plane.  One can see here the difference to the usual Bloch vector representation given in the diabatic basis. There the field vector associated with the model in Eq.~(\ref{eqn:superparabolic}) is given as $\tilde{\vec{B}} = [\Omega_{0}, 0, \beta^{N+1}t^{N}]^{T}$, so initially ($t = -\infty$), it is aligned with the $z$-axis and its length is infinite. It is also confined in the $x$-$z$ plane at all times. If the system is initially prepared to an eigenstate of the Hamiltonian, that is $z(-\infty) = \pm 1$, then the Bloch vector will at first follow adiabatically the field vector but will eventually start to deviate and to precess around it as the direction of $\vec{B}$ changes. In the rotating frame the same initial conditions mean that the Bloch vector and the field vector are initially orthogonal and the adiabaticity can be understood from the fact that when the system is far away from the resonance point $t = 0$, the field vector is rotating so rapidly that its effects are averaged out as can be seen the particular choice of $\phi(t)$ in Eq.~(\ref{eqn:windinghamiltonian}). The difference between the detuning functions of the superparabolic level glancing models and the cubic-like level crossing models in \cite{vitanov} is reflected to the behaviour of the field vectors, so that, in the rotating frame, the field vector will reverse its direction of rotation at $t=0$ for superparabolic models whereas for supercubic models it does not.

By taking the initial conditions to be $z(-\infty) = -1$, $y(-\infty) = x(-\infty)= 0$ and integrating the equations of motion (\ref{eqn:xyzderivative}) for the Bloch vector components, we get an expression for the time evolution of the $z$-component as
\begin{widetext}
\begin{equation}
\begin{aligned}
z(t) &= -1 - 4\int_{-\infty}^{t} d t^{'} \Omega(t^{'})\lbrace \cos [ 2\phi( t^{'} ) ] \int_{-\infty}^{t^{'}} d t^{''}  \Omega(t^{''})\cos [ 2\phi( t^{''} ) ] z(t^{''}) +  \sin [ 2\phi( t^{'} ) ] \int_{-\infty}^{t^{'}} d t^{''} \Omega(t^{''})\sin [ 2\phi( t^{''} ) ] z(t^{''})\rbrace  \\
&= - 1 - 4\int_{-\infty}^{t} d t^{'} \int_{-\infty}^{t^{'}} d t^{''} \Omega(t^{'})\Omega(t^{''})\cos [ 2\phi( t^{'} ) - 2\phi( t^{''} )] z(t^{''}) .
\end{aligned}\label{eqn:z}
\end{equation}
But we can also consider the one step before Eq.~(\ref{eqn:z}), namely
\begin{equation}
\dot{z}(t) = - 4\Omega(t) \lbrace \cos \left[ 2\phi( t ) \right] \int_{-\infty}^{t} d t^{'} \Omega(t^{'})\cos \left[ 2\phi ( t^{'} ) \right] z(t^{'}) +  \sin [ 2\phi( t ) ] \int_{-\infty}^{t} d t^{'} \Omega(t^{'})\sin [ 2\phi( t^{'} ) ] z(t^{'})\rbrace .  \label{eqn:zderiv}
\end{equation}
\end{widetext}
Multiplying both sides of Eq.~(\ref{eqn:z}) by $z(t)$ and defining the functions
\begin{equation}
\begin{aligned}
F_{c}(t) &= \int_{-\infty}^{t} d t^{'} \Omega(t^{'})\cos [ 2\phi( t^{'} ) ] z(t^{'}),\\
F_{s}(t) &= \int_{-\infty}^{t} d t^{'} \Omega(t^{'})\sin [ 2\phi( t^{'} ) ] z(t^{'}), 
\end{aligned}
\end{equation}
we get the form
\begin{equation}
z(t)\dot{z}(t) = - 4 \lbrace F_{c}(t)\dot{F}_{c}(t) +  F_{s}(t)\dot{F}_{s}(t) \rbrace.\label{eqn:zdotz}
\end{equation}
By integrating Eq.~(\ref{eqn:zdotz}) from $-\infty$ to $t$ and imposing the initial condition $z^{2}(-\infty) = 1$ we get
\begin{equation}
z^{2}(t) = 1 -  4\lbrace F_{c}^{2}(t) + F_{s}^{2}(t) \rbrace.
\label{eqn:inteqforsquarepop}
\end{equation}
Let us now explicitly consider the superparabolic models. To this end, it is useful to first scale the time as $\tau = \beta t $, leaving all the parameter dependence of the model Hamiltonian given in the diabatic representation by Eq.~(\ref{eqn:superparabolic}) on the off-diagonal elements. We further define this new diabatic coupling as 
\begin{equation}
\alpha = \Omega_{0} /  \beta \label{eqn:alphadef}
\end{equation} 
so that the limit $\alpha \rightarrow \infty$ is now the adiabatic limit and $\alpha \rightarrow 0$ is the diabatic limit. By choosing the time-dependence of the transformation to the rotating frame as $\phi(\tau) = \tau^{N+1}/(N + 1)$, we get Eq.~(\ref{eqn:inteqforsquarepop}) into the form 
\begin{widetext}
\begin{equation}
z^{2}(\tau) = 1 - \alpha^{2} \left[ 2^{N} (N + 1)\right]^{2/(N + 1)} \left[ \left( \int_{-\infty}^{\tau} ds \cos (s^{N + 1} ) z(s) \right)^{2} +  \left( \int_{-\infty}^{\tau} ds \sin (s^{N + 1} ) z(s) \right)^{2} \right].
\end{equation}
In the limit of weak coupling, this simplifies to 
\begin{equation}
z^{2}(\tau) = 1 - \alpha^{2} \left[ 2^{N} (N + 1)\right]^{2/(N + 1)} \left[ \left( C_{N + 1}(\tau) + C_{N + 1}(\infty)\right)^{2} + \left( S_{N + 1}(\tau) - S_{N + 1}(\infty)\right)^{2}\right],
\label{eqn:weakc}
\end{equation}
\end{widetext}
where 
\begin{equation}
\begin{aligned}
C_{n}(\tau) &= \int_{0}^{\tau} ds \cos (s^{n} ), \\
S_{n}(\tau) &= \int_{0}^{\tau} ds \sin (s^{n} ),
\end{aligned}
\label{eqn:genfrsnlint}
\end{equation}
are the generalized Fresnel integrals \cite{genfresnel}, defined for $\tau \geq 0$ and $ n > 1$. The limits of the integrals are given by
\begin{equation}
\begin{aligned}
C_{n}(\infty) &= \frac{\Gamma\left(1/n\right)}{n} \cos\left(\pi / 2 n \right),\\
S_{n}(\infty) &= \frac{\Gamma\left(1/n\right)}{n} \sin\left(\pi / 2 n \right),
\label{eqn:genfrsnlintlimits}
\end{aligned}
\end{equation}
so, in the weak coupling case, we get
\begin{widetext}
\begin{equation}
z^{2}(\infty) = 1 - 4\alpha^{2} \left( \dfrac{2}{N + 1} \right)^{2N/(N + 1)} \Gamma^{2}\left( \dfrac{1}{N + 1}\right) \cos^{2}\left( \dfrac{\pi}{2(N+1)}\right).
\end{equation}
\end{widetext}
Furthermore, as the assumption of weak coupling means that the diabatic coupling is too weak to change the populations considerably, it follows from the initial condition $z(-\infty) = -1$ that $z(\tau) \approx -1$ for all time, so we can approximate $z^{2}(\infty) - 1 \approx -2(z(\infty) + 1)$. Now, from the definition of the Bloch vector in Eq.~(\ref{eqn:xyz}) we also have $P = (1 + z(\infty))/2$, so that we finally get an expression for the final transition probability in the weak coupling approximation as
\begin{widetext}
\begin{equation}
P_{pert} = \alpha^{2} \left( \dfrac{2}{N + 1} \right)^{2N/(N + 1)} \Gamma^{2}\left( \dfrac{1}{N + 1}\right) \cos^{2}\left( \dfrac{\pi}{2(N+1)}\right).
\label{eqn:perturbationp}
\end{equation}
\end{widetext}
This approximation is expected to give accurate results only for very small values of the diabatic coupling and, indeed, it gives values over unity already with parameter values of $\alpha \approx 1/2$. To expand the range of validity of the perturbative approximation and to take the oscillatory character of the final transition probability into account one can also consider the Magnus approximation \cite{light, kasphd} which can be obtained directly from Eq.~(\ref{eqn:perturbationp}) with the formula
\begin{equation}
P_{Magnus}= \sin^{2}\left(\sqrt{P_{Born}}\right), 
\end{equation}
which, of course, is always less than or equal to unity. Although this is an improvement for the models with larger $N$, in general this approximation is only of limited value. These approximations, along with the DDP approximation which is derived in the next section, are compared to numerical results.

\subsection{Dykhne-Davis-Pechukas formula}

\begin{figure}
\begin{center} 
\includegraphics[scale=1.0]{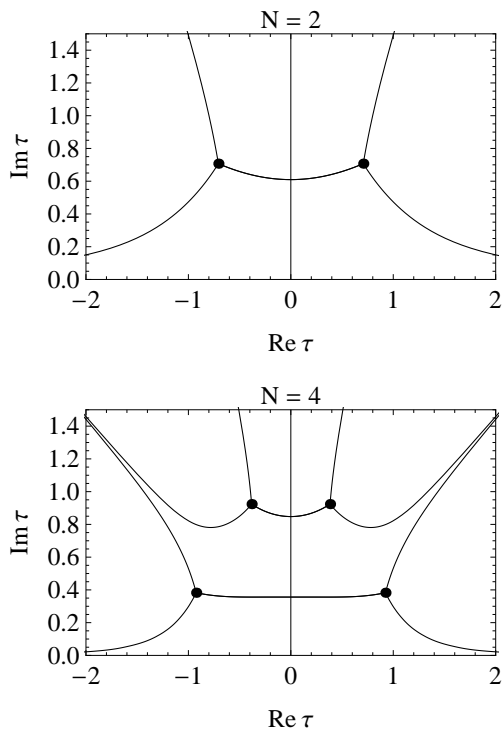} 
\caption{The complex zero points of the quasienergies are shown as black dots and the structure for the Stokes lines as solid lines for superparabolic models $N = 2$ and $N = 4$. We have chosen $\alpha = 1$ so that the zero points lie on unit circle. The corresponding figures for higher values of $N$ are similar, and in general there are three Stokes lines emanating from each zero point, and the points with equal imaginary part are linked by a Stokes line.}
\label{stokesinviivat}
\end{center}
\end{figure}

The Dykhne-Davis-Pechukas (DDP) formula is  given by
\begin{equation}
P = e^{- 2 Im D(t_{c})},
\label{eqn:dykhne}
\end{equation}
where
\begin{equation}
D(t) = 2 \int_{0}^{t} \mathcal{E}(s) ds,
\label{eqn:D}
\end{equation}
and $t_{c}$ is defined by the equation
\begin{equation}
\mathcal{E}(t_{c}) = 0.
\end{equation}
Equation (\ref{eqn:dykhne}) gives the probability of nonadiabatic transitions that is asymptotically exact in the adiabatic limit. This was proven by Davis and Pechukas \cite{dp}, who followed the original idea of Dykhne \cite{dykhne}. The main assumptions behind this formula are that $\mathcal{E}(t)$ is non-vanishing for real $t$ (including $t = \pm \infty$), that $t_{c}$ is well separated from other zero points or possible singularities and that the Hamiltonian is analytic and single-valued at least in a region of complex $t$ plane bounded by the real axis and the Stokes lines nearest to the real axis. The Stokes lines are defined as the level lines of $D(t)$ with $D(t) = D(t_{c})$. Assuming that the zero points of $\mathcal{E}(t)^{2}$ are simple, it is easy to see by local analysis that there are three lines emanating from $t_{c}$ with equal angles to each other as shown in Fig.~\ref{stokesinviivat}. The zero points come in conjugate pairs, but we can restrict our considerations to upper half plane where $\mathfrak{Im}(t) > 0$.
  
The DDP formula can generally be expected to a give a good approximation for $P$ only near the adiabatic limit but for example for LZ model it gives the exact result, so it is correct for all the parameters. In the LZ case the only zero point on the upper half of the complex plane lies on the imaginary axis and Eq.~(\ref{eqn:dykhne}) gives  
\begin{equation}
P = e^{-\pi \alpha^{2}}.
\end{equation}
In the case that there exists more than one complex zero point, as is the case for the superparabolic models defined by Eq.~(\ref{eqn:superparabolic}), the DDP formula (\ref{eqn:dykhne}) can be directly generalized to include the contributions of all the complex zeros relevant to the problem, as suggested in Refs.~\cite{dp,kasphd,george}, so that 
\begin{equation}
P = \left\vert \sum_{k = 1}^{N} \Gamma_{k} e^{i D\left(t_{c}^{k}\right)} \right\vert^{2}, 
\label{eqn:genDDP}
\end{equation}
where
\begin{equation}
\Gamma_{k} = 4 i \lim_{t \rightarrow t_{c}^{k}} \left( t - t_{c}^{k}\right) \gamma(t),
\label{eqn:isogamma}
\end{equation}
and $\gamma(t)$ is the nonadiabatic coupling defined in Eq.~(\ref{eqn:nonadiabcoupl}). From this formula it can be seen that the existence of multiple zero points lead to oscillations in the final state populations as the parameters are varied. 

The obvious problem in this case is to decide which of the points should be taken into account. As was shown rigorously by Joye and coworkers in Refs.~\cite{joye, joye2}, this question is related to the global structure of the set of Stokes lines and the correct way is to include only the points connected by the Stokes line nearest to the real axis. However, they used an assumption that a limiting Hamiltonian exists as $t\rightarrow \pm \infty$ which is not valid for many interesting models, including the present superparabolic models. Instead, it may be beneficial to include all the zero points in the $\mathfrak{Im}(t) > 0$ plane in Eq.~(\ref{eqn:genDDP}) as studied in \cite{kasphd,vitanov} and this is also the possibility we consider here.  This viewpoint is supported by the fact that for the Demkov-Kunike models the sum over an infinite set of zero points actually produces the known exact result for all parameters~\cite{kasphd}.

\subsubsection{Application to superparabolic models}

The zero points of the eigenvalues of the superparabolic Hamiltonian are
\begin{equation}
\tau_{c}^{k} = \alpha^{1/N} e^{i \pi (2 k - 1)/ 2N}, \; k = 1, 2, \ldots, N,
\end{equation}
so the zero points lie on a circle of radius $\alpha^{1/N}$. An example of the zero points and the structure of the Stokes lines is given in the Fig.~\ref{stokesinviivat}. We get from Eq.~(\ref{eqn:D})
\begin{equation}
D(\tau_{c}^{k}) = \eta e^{i \pi (2k - 1)/2N},
\end{equation}
where
\begin{equation}
\eta = 2 \nu_{N}\alpha^{(N + 1)/N}
\end{equation}
and
\begin{equation}
\nu_{N} = \int_{0}^{1} \sqrt{1 - y^{2N}} dy = \frac{1}{2N} B\left( \frac{1}{2N}, \frac{3}{2}\right),
\end{equation}
where $B(x, y)$ is the beta function~\cite{NIST}. The function $\nu_{N}$ tends to unity as $N$ increases. The factors in Eq.~(\ref{eqn:isogamma}) are given by $ \Gamma_{k} = (-1)^k$ and the points $\tau_{c}^{k}$ can be grouped into pairs with the same imaginary part and opposite real part, so that the generalized DDP formula for the superparabolic models can be written in the form
\begin{widetext}
\begin{equation}
P_{DDP} = 4 \left\vert \Sigma_{k=1}^{N/2} (-1)^{k} e^{- \eta \sin \left[\frac{\pi}{2N}(2k - 1)\right]} \sin \left[\eta \cos{\frac{\pi}{2N} (2k - 1)} \right] \right\vert^{2}.
\label{eqn:ddpsuperparabolinen}
\end{equation}
In many ways this is similar as the corresponding formula for the model in Eq.~(\ref{eqn:superparabolic}) with odd $N$ obtained in Ref.~\cite{vitanov},
\begin{equation}
P_{DDP}^{odd} = 4 \left\vert \Sigma_{k=1}^{(N - 1)/2} (-1)^{k} e^{- \eta \sin \left[\frac{\pi}{2N}(2k - 1)\right]} \cos \left[\eta \cos{\frac{\pi}{2N} (2k - 1)} \right] + \left( -1 \right)^{(N+1)/2} e^{-\eta}  \right\vert^{2},
\label{eqn:ddpnonlinear}
\end{equation}
\end{widetext}
except that with even $N$ the purely imaginary zero point is missing, which gives the last term in Eq.~(\ref{eqn:ddpnonlinear}). Such a purely imaginary zero point is present in the LZ model as the only zero point, but in Eq.~(\ref{eqn:ddpnonlinear}) the largest contribution in the large-$\alpha $ limit comes in fact from the first term of the sum and the role of the LZ-like term is suppressed. Thus the values for $P$ show oscillations as a function of $\alpha$ also for odd values of $N$, as long as $N>1$. This is a simple mathematical explanation why the transition probabilities show oscillations that one would expect mainly for models that have interferometric character, i.e., two or more actual crossing or glancing points in the diabatic basis. There is no obvious physical reason for such behaviour, although for even values of $N$ one can understand the level glancing situation as a limiting case for the more general superparabolic model with double crossings~\cite{suominen2,kasphd}.

\section{Comparison of methods}\label{sec:results}

\begin{figure}[htb]
\begin{center}
\includegraphics[scale=0.8]{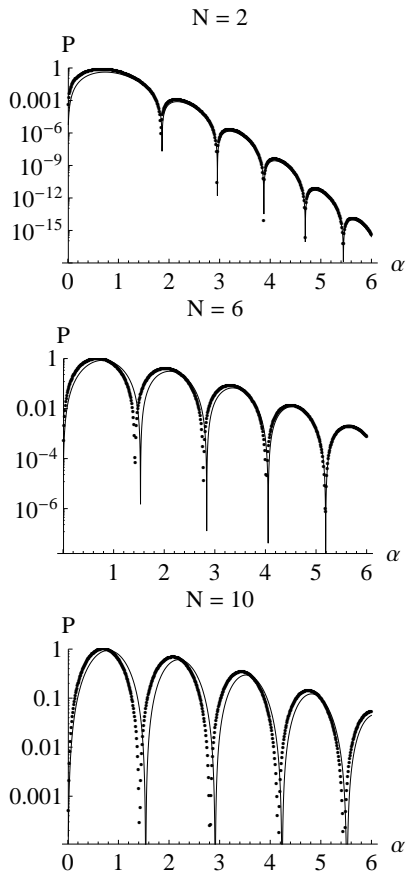} 
\caption{The probability $P$ of a nonadiabatic transition is plotted as a function of the coupling $\alpha$ for the values $N = 2, \, 6$ and 10. The dots indicate the numerical solution and the solid line is the generalized DDP result.}
\label{poplog}
\end{center}
\end{figure}

\begin{figure}[htb]
\begin{center} 
\includegraphics[scale=0.67]{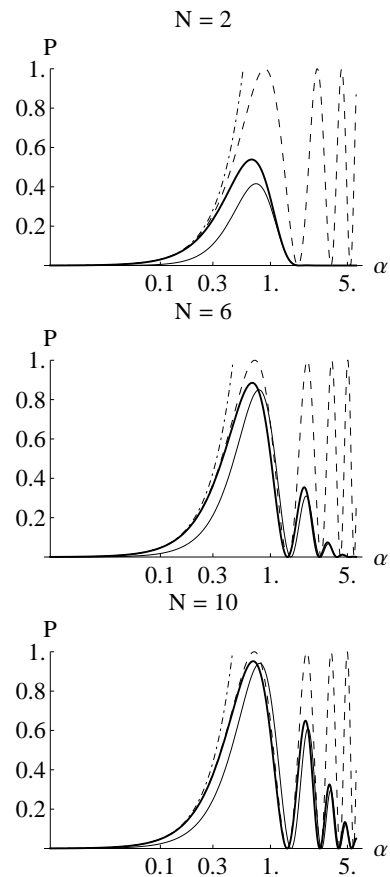} 
\caption{This figure is similar to Fig.~(\ref{poplog}) but the vertical scale is normal whereas the horizontal scale is logarithmic. In this figure, the thick solid line is the numerical result and the thin line is the DDP result. Furthermore, the dot-dashed line in this figure represents the weak coupling approximation $P_{Pert}$ and the dashed line is the Magnus approximation.}
\label{pop2610}
\end{center}
\end{figure}

We have applied the perturbative methods, the DDP methods and numerical evaluation to the superparabolic level glancing models. The asymptotic probabilities $P_{DDP}$ and $P_{Magnus}$ for the parabolic and superparabolic models are shown in Figs.~\ref{poplog} and \ref{pop2610} along with the results obtained from numerical calculations as the coupling $\alpha$ is varied from 0 to 6. They are oscillating functions with respect to $\alpha$ in contrast to the monotonous relationship between the probability of transition in the LZ model and its adiabatic parameter. In Fig.~\ref{pop2610} we show the same curves but with logarithmic scale for the coupling in order to study the approximations near the maximum values of the transition probability. We demonstrate the divergence of the weak coupling approximation $P_{Pert}$ in this figure.  But before making the full comparison, let us discuss a few special points in light of the shown results.

\begin{figure}[htb]
\begin{center} 
\includegraphics[scale=0.7]{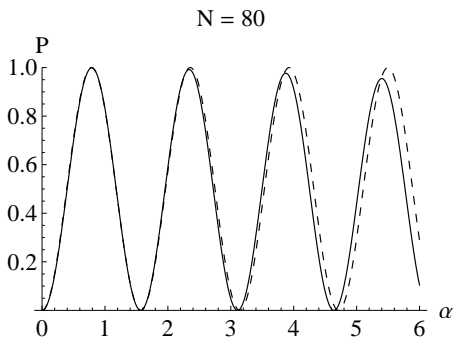} 
\caption{This figure demonstrates the behaviour of $P_{DDP}$ (solid line) and $P_{Rabi}$ (dashed) for large $N$. Here we have chosen $N = 80$. }
\label{rabiN80}
\end{center}
\end{figure}

One can see that the validity of the generalized DDP approximation extends to increasingly smaller values of $ \alpha $ as $ N $ grows and apart from small phase shift it is in a good agreement with the numerical result for all values of $\alpha$. Let us next consider the behaviour of the superparabolic models as $ N \rightarrow \infty $. As already mentioned, the behaviour of the diabatic energy levels in the limit of large $N$ indicates that one could expect the time-evolution of the system to be divided into two parts, i.e., to a non-resonant part where the time-evolution is not affected by the coupling, and to a resonant period where the dynamics is described by the well-known Rabi model. The Rabi model corresponding to the superparabolic model in this resonant period consists of a constant rectangular pulse, of magnitude $ \alpha $, which couples the levels. Thus one expects to obtain the final transition probability $P$ by just determining the duration of the resonant period and inserting it to the well-know expression for resonant Rabi oscillations. To get an estimate for this duration time we use the same argument that is generally used e.g. in estimating the dynamically relevant time region for the LZ model (this crude approximation is sufficient if $\alpha$ is not too large, see Ref.~\cite{Vitanov1999}). Then the start and the end of the transition region is determined simply from the condition $\Delta(\tau) = \Omega(\tau)$,  which gives $\Delta \tau = 2 \alpha^{1/N}$, so that in the limit $ N \rightarrow \infty$ the final transition probability according to the Rabi formula is
\begin{equation}
P_{Rabi} = \sin^{2} \left[ 2 \alpha \right].
\label{eqn:rabi}
\end{equation}
This is consistent with the results of the DDP approximation with large $N$ as can be seen from Fig.~\ref{rabiN80}. As expected, deviations begin to occur when $\alpha$ increases as then our crude estimate for resonant period begins to falter.

To evaluate the difference between the generalized DDP approach and the original DDP theory, we consider $P_{DDP}$ when one includes in the sum only the zero points closest to the real axis. We denote this original theory prediction by $P_{DDP}^{(1)}$. As it has been shown there are two such points located symmetrically with respect to imaginary axis and those correspond to the first term in the summation in Eq.~(\ref{eqn:ddpsuperparabolinen}), so that 
\begin{equation}
P_{DDP}^{(1)} = 4 e^{-2 \eta \sin \left[\frac{\pi}{2N}\right]} \sin^{2} \left[\eta \cos{\frac{\pi}{2N} } \right].
\end{equation}
In the case of weak coupling, $\alpha \approx 0$, this becomes
\begin{equation}
P_{DDP}^{(1)} \approx 4 \pi \alpha^{(2N + 1)/N} \frac{\Gamma^{2} \left[ (2N + 1)/(2N) \right]}{\Gamma^{2} \left[ (3N + 1)/(2N) \right]}  \cos^{2}\left(\frac{\pi}{2N}\right), 
\end{equation}
which is of interesting form when compared to the proper weak coupling limit in Eq.~(\ref{eqn:perturbationp}). When $N$ increases they both tend to limit $\alpha \rightarrow 0$ as $\alpha^{2}$, but for finite $N$ the two expressions are slightly different. For the generalized DDP result in Eq.~(\ref{eqn:ddpsuperparabolinen}), we get a sum with all terms proportional to $\alpha^2$ in the small $\alpha$ limit, but the sum does not appear to have any simple closed form for easy comparison.

\begin{figure}[tb]
\begin{center}
\includegraphics[scale=0.7]{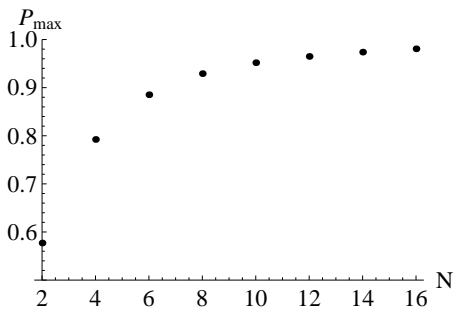} 
\caption{The maximum probabilities that can be obtained for each model as a function of $N$.
\label{pmaksimit}}
\end{center}
\end{figure}

Let us now proceed with the general examinations of the results. We can see from the Fig.~\ref{pmaksimit} and Table~\ref{taulukko} that the maximum probability is, as we expect, enhanced by increasing $N$, though the rate of the increment decreases steadily. At the same time, the value of coupling $\alpha_{0}$ needed to obtain the maximum value increases only slowly.

\begin{table}[tb]
\begin{ruledtabular}
\begin{tabular*}{\hsize}{r@{\extracolsep{0ptplus1fil}}c@{\extracolsep{0ptplus1fil}}l}
$N$&$P_{max}$&$\alpha_{0}$\\
\colrule
2&0.577&0.68\\
4&0.792&0.68\\
6&0.885&0.68\\
8&0.929&0.69\\
10&0.952&0.70\\
12&0.965&0.70\\
14&0.974&0.71\\
16&0.981&0.71\\
\end{tabular*}
\end{ruledtabular}
\caption{The maximum values of $P$ that appear in Fig.~\ref{pmaksimit} and also the values of coupling $\alpha_{0}$ with which they can be obtained.
\label{taulukko}}
\end{table}

It seems that the parameter value $\alpha_{0}$ for which the maximum transition probability is obtained is somewhere in between the region of validity of the perturbation and DDP approximations. It is, however, evident from Figs.~\ref{poplog} and \ref{pop2610} that the generalized DDP method is in good agreement with the numerical results for the superparabolic models in a large part of the parameter region and not just in the adiabatic limit, and its accuracy gets better as $N$ increases. Because of the oscillating character of the transition probability, small error in the phases can lead to notable deviation from the true value, and there indeed is some phase difference in the DDP and numerical results but the generalized DDP results still catches all the essentials of the numerical results. For small values of $\alpha$, the perturbation approximation is more accurate, its region of validity remaining somewhat constant. The Magnus approximation improves the perturbation result further but in a limited fashion.

It is clear that to get a quantitatively accurate approximation using the generalized DDP method, one has to include all complex zero points into the expression (\ref{eqn:genDDP}). Taking into account only the pair of complex zero points closest to the real axis and which are connected by the closest Stokes line gives qualitatively somewhat correct behaviour and for example the parameter value for obtaining the maximum probability but also values over unity.     

\section{Conclusions}
\label{sec:conclusions}

Although the Landau-Zener model has become, for many reasons, a paradigm in studies of coupled time dependent quantum states, the parabolic model and especially the level glancing aspect have not been considered in detail, and the superparabolic models have not been considered at all, to the best of our knowledge. A level glancing situation is, of course, much less likely to occur in nature, although increasing interest and development of tools for control and engineering of quantum states will make it likely that such models can be tested and applied in experimental physics. Our motivation for the present study has been three-folded.

Firstly, we wanted to understand the nature of the level glancing dynamics and especially the oscillating character of the transition probabilities, i.e., how the properties of the parabolic model extend to the superparabolic models. We have shown that the superparabolic models provide dynamics that follows qualitatively the parabolic model, and eventually starts to resemble the Rabi model, with a sudden approach to the resonance, steady resonant Rabi oscillations, and then the freezing of the value of the transition probability to the moment of the sudden move away from the resonance.

Secondly, we have complemented the picture of essentially nonlinear models as studied in Ref.~\cite{vitanov}, by introducing the superparabolic models, and showing how the structure of the complex zeros reflects to the transition probability. It is clear that the Landau-Zener model is in many ways unique in its monotonous change of transition probability as a function of coupling strength, whereas both the cubic-like and the superparabolic models will display oscillations, and the superparabolic models allow one to reach full transition only asymptotically. 

Thirdly, the improvement of the adiabatic solution (DDP) by addition of all zeros of the eigenenergies in the complex plane lacks rigorous proof so far, and its practicality must be evaluated for each model independently~\cite{kasphd}. Our work shows that for the superparabolic level glancing models the inclusion of all such zeros in the upper half of the complex plane improves the quality of the solution and extends its usefulness beyond the adiabatic limit. 

One aspect of the Landau-Zener model is the intrinsic phase difference between the quantum state amplitudes which the level crossing transition dynamics provides~\cite{kazantsev}. This becomes visible in transition probabilities for any double or multiple crossings situation, although the dynamical phase evolution of the amplitudes between the crossings tends to dominate any oscillations~\cite{suominen2}. The other models, including parabolic and superparabolic models, give rise to a similar phase factor. The Landau-Zener model has been applied to periodically driven systems~\cite{Garraway1992,Foldi2008,Cao2010}, and it would be interesting to see if a system with periodically occurring level glancing events gives similar dynamics. A second aspect to consider is the role of noise on both the cubic-like as well as parabolic and superparabolic models. The Landau-Zener model as such is not very much affected by the noise~\cite{Akulin1992}, but any double or multiple crossing situation will be strongly affected by phase degradation~\cite{sillanpaa}, offering, on the other hand, a tool for analysing decoherence of quantum states. Control of two-state systems as well as their decoherence is important for quantum information and quantum computing~\cite{Nielsen2000}. The essentially nonlinear models all show oscillations in the transition probabilities, even though there is only a single level crossing or a level glancing event. Their sensitivity to decoherence is an interesting and open question, and one of the topics for further studies on level glancing models.

\begin{acknowledgments}
This research was supported by the Finnish Academy of Science and Letters, and the Academy of Finland, grant 133682. The authors thank A. Ishkhanyan for discussions.
\end{acknowledgments}

\end{document}